\documentclass[prb,aps,twocolumn,amssymb,showpacs]{revtex4}
\usepackage{graphicx}
\usepackage{longtable}
\usepackage{epsfig}
\usepackage{dcolumn}
\usepackage{bm}

\begin{document}

\title{       Origin of the Verwey transition in magnetite: \\
              Group theory, electronic structure and lattice dynamics study}

\author{      Przemys\l{}aw Piekarz }
\affiliation{ Institute of Nuclear Physics, Polish Academy of Sciences, 
              Radzikowskiego 152, PL-31342 Krak\'{o}w, Poland }

\author{      Krzysztof Parlinski }
\affiliation{ Institute of Nuclear Physics, Polish Academy of Sciences, 
              Radzikowskiego 152, PL-31342 Krak\'{o}w, Poland }

\author{      Andrzej M. Ole\'{s} }
\affiliation{ Institute of Nuclear Physics, Polish Academy of Sciences, 
              Radzikowskiego 152, PL-31342 Krak\'{o}w, Poland }

\begin{abstract}
The Verwey phase transition in magnetite has been analyzed using the 
group theory methods. It is found that two order parameters with the 
symmetries $X_3$ and $\Delta_5$ induce the structural transformation 
from the high-temperature cubic to the low-temperature monoclinic phase.
The coupling between the order parameters is described by the Landau 
free energy functional. The electronic and crystal structure for the 
cubic and monoclinic phases were optimized using the {\it ab initio\/} 
density functional method. The electronic structure calculations were 
performed within the generalized gradient approximation including 
the on-site interactions between $3d$ electrons at iron ions --- the 
Coulomb element $U$ and Hund's exchange $J$. Only when these local 
interactions are taken into account, the phonon dispersion curves, 
obtained by the direct method for the cubic phase, reproduce the 
experimental data. It is shown that the interplay of local electron 
interations and the coupling to the lattice drives the phonon order 
parameters and is responsible for the opening of the gap at the Fermi 
energy. Thus, it is found that the metal-insulator transition in 
magnetite is promoted by local electron interactions, which 
significantly amplify the electron-phonon interaction and stabilize 
weak charge order coexisting with orbital order of the occupied 
$t_{2g}$ states at Fe ions. This provides a scenario to understand the 
fundamental problem of the origin of the Verwey transition in magnetite.
\end{abstract}

\date{\today}


\maketitle

\section{Introduction}

For many decades, magnetite --- the oldest known magnetic material --- 
has been a fascinating subject of extensive research.
Besides its distinct magnetic properties, the main interest was focused
on the mechanism and physical consequences of the first-order phase 
transition at $T_V=122$ K, called the Verwey transition (VT), 
in which conductivity changes by about two orders of magnitute.
\cite{verwey1} For a long time it has been considered as an 
example of charge localization driven by metal-insulator transition,
\cite{verwey2} in which ionic interactions determine the electronic 
properties.\cite{anderson} Only in the last decade the progress 
in experimental and theoretical techniques, as well as improved 
quality of samples, allowed one for a deeper insight into the nature 
of this transition.\cite{review} Recent studies demonstrated that the 
VT is a cooperative phenomenon, in which an interplay between lattice, 
charge, and orbital degrees of freedom plays a decisive role. In this 
context, magnetite remains to be one of the most interesting materials 
among transition metal oxides.\cite{imada}

\begin{figure}[b!]
\includegraphics[width=7.9cm]{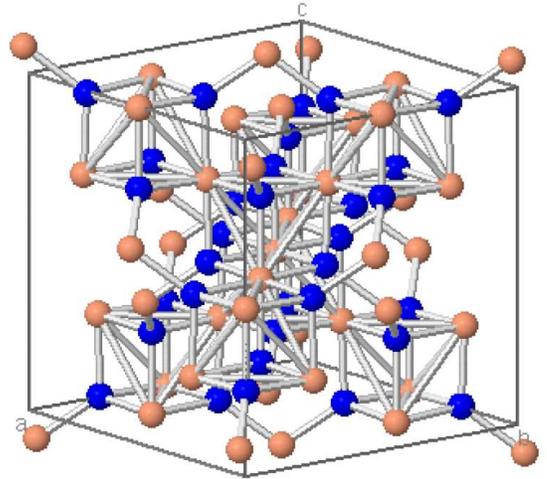}
\caption{(Color online)
The crystal structure of magnetite in the cubic $Fd\bar{3}m$ symmetry.
Iron and oxygen ions are represented by grey and dark (orange and blue)
balls. Fe($B$) ions have six O neighbors, while each O ion has three
Fe($B$) and one Fe($A$) neighbor.
}
\label{fig:fig1}
\end{figure}

At room temperature, magnetite (Fe$_3$O$_4$) crystalizes in the inverse 
spinel cubic structure with Fe ions occupying the tetrahedral $A$ sites 
and octahedral $B$ sites, as shown in Fig. \ref{fig:fig1}. The ionic 
structure of magnetite in the inverse spinel phase is 
Fe$^{3+}(A)$Fe$^{2.5+}(B)$Fe$^{2.5+}(B)$O$_4^{2-}$. Below the N\'eel 
temperature $T_N=851$ K, magnetic moments at Fe ions are aligned 
antiparallely between $A$ and $B$ sites, which results in the 
ferrimagnetic state. Verwey proposed that the metal-insulator transition 
is caused by the charge order (CO) of Fe$^{2+}$ and Fe$^{3+}$ ions at 
the $B$ sites, in planes perpendicular to the $c$ axis.\cite{verwey2} 
The structural analysis, however, revealed the crystal distortion, which 
is incompatible with the Verwey model. In particular, the observation of 
half-integer reflections at $(h,k,l+\frac{1}{2})$ points indicated the
doubling of the unit cell along the $c$ axis.\cite{struct1} Further 
diffraction studies by neutron,\cite{struct2,struct3} x-ray,\cite{struct4}
and electron \cite{struct5} scattering established the monoclinic symmetry
of the low temperature (LT) phase realized below $T_V$. Indeed, the number 
of inequivalent $A$ and $B$ atomic positions found by nuclear magnetic 
resonance (NMR) for the phase below the VT
\cite{NMR1,NMR2,NMR3} agrees with the monoclinic structure.

The models to describe the CO in magnetite, consistent with the LT 
phase of Fe$_3$O$_4$, have been proposed in many studies.
\cite{struct5,NMR1,NMR2,NMR3} High-resolution neutron and x-ray 
scattering measurements\cite{attfield1,attfield2} revealed the 
distribution of Fe-O bond distances in the $B$O$_6$ octahedra (between 
1.96 and 2.11 \AA), being the indication of charge disproportionation 
on Fe($B$) ions. The CO proposed by Attfield {\it et al.}
\cite{attfield1,attfield2} is considerably more complex than the Verwey 
model\cite{verwey1} itself, and consists of four different $B$ sites 
in a unit cell, with approximately two valence states: Fe$^{+2.4}$ and 
Fe$^{+2.6}$. Providing a convincing evidence in favor of such a state 
has been a challenge for several direct probing methods including the 
resonant x-ray scattering. Despite preliminary negative results,
\cite{CO1,CO2} the evidence that the CO is indeed fractional has 
accumulated recently.\cite{CO3,CO4,CO5,CO6} It is clear from the 
character of the observed CO that a simple ionic mechanism suggested 
early on by Anderson\cite{anderson} cannot explain the VT.

The electronic structure of magnetite was studied by the density functional 
theory (DFT) calculations in the local spin density approximation (LSDA).
\cite{DFT1} This approach is based on the local density approximation (LDA)
and provides a basic description of the electronic structure and magnetic 
properties of Fe$_3$O$_4$ in the cubic phase above $T_V$.
\cite{DFT1,DFT2,DFT3} The CO instability induced by local on-site 
and intersite Coulomb interactions was studied in the cubic symmetry 
within the LSDA+$U$ method.\cite{LDAU1,LDAU2} For a distorted (orthorhombic) 
symmetry, the self-interaction correction\cite{szotek} approach was used, 
and it has been shown that the CO proposed by Verwey does not appear in the 
ground state of magnetite. The calculations for the experimentally observed 
monoclinic structure, performed within the LSDA+$U$ (see Refs. 
\onlinecite{madsen,prl1}) and generalized gradient approximation (GGA) 
with local Coulomb interaction $U$ (GGA+$U$, see, e.g. Ref. 
\onlinecite{prl2}) approaches proved the existence of fractional charge 
disproportionation in fair agreement with the experimental analysis. 
\cite{attfield1} In addition, these studies revealed the presence of 
the orbital order (OO) of the occupied $t_{2g}$ states at $B$-sites,
\cite{prl1,prl2} in agreement with recent experiments performed on thin 
films of Fe$_3$O$_4$.\cite{Nea07} The obtained magnitude of the insulating 
gap agrees well with the photoemision data below $T_V$.\cite{PES1,PES2,PES3} 
On the contrary, in absence of electron interactions (for $U=0$) the 
electronic state remains metalic even in the distorted structure, 
showing the essential role played by local electron interactions in 
the VT. These {\it ab initio\/} calculations demonstrate also that the 
microscopic understanding of the VT requires to involve 
lattice distortions and electron-phonon (EP) interaction. 

The experimental evidence that phonons participate in the mechanism of 
the VT is very convincing. The structural analysis showed that the 
LT phase results from the condensation of modes at three reciprocal 
lattice points: $\textbf{k}_{\Gamma}=(0,0,0)$, 
$\textbf{k}_{\Delta}=(0,0,\frac{1}{2})$, and $\textbf{k}_X=(0,0,1)$, 
\cite{struct3,attfield2} in units of $\frac{2\pi}{a}$, where $a$ is 
lattice constant of the cubic structure. The pronounced softening of 
the $c_{44}$ elastic constant,\cite{elastic1} observed from room 
temperature down to $T_V$, has been explained by the coupling between 
the shear mode and the charge-density $T_{2g}$ order parameter (OP) at 
the $\Gamma$ point.\cite{elastic2} The decrease of $c_{44}$ is 
associated with the softening of the surface Rayleigh mode observed by 
the Brillouin scattering.\cite{brillouin} In the bulk, no phonon 
softening has been observed,\cite{phonons} however, critical 
fluctuations revealed by neutron scattering strongly indicate the 
existence of phonon precursor effects which drive the structural 
deformation. First, the spotlike diffuse 
signal was observed at the reciprocal space point 
${\bf k}_{\Delta}=(4,0,\frac{1}{2})$.\cite{diff1} Yamada proposed that 
the VT results from the condensation of a coupled charge density-phonon 
mode with $\Delta_5$ symmetry, explaining the intensities of neutron 
signal and $\textbf{k}_{\Delta}$ Bragg reflections.\cite{yamada1}
Second, a new type of planar diffuse scattering was found over a broad 
region of the reciprocal space at high temperature $T_V<T<T_V+100$ K.
\cite{chiba, diff2} The signal around the $\Gamma$ and equivalent 
points was interpreted as the Huang scattering due to the local strain 
field, and was studied in the molecular polaron model.
\cite{diff3,yamada2} The analysis of intensities at $\textbf{k}_X$ 
points led to the conclusion that transverse phonons with the $X_3$ 
symmetry dominate in neutron scattering.\cite{diff5}
The phonon mechanism is further supported by oxygen isotope effect, 
\cite{isotope} Raman,\cite{Raman1,Raman2,Raman3} the extended x-ray 
absorption fine structure (EXAFS),\cite{EXAFS} 
and by nuclear inelastic scattering (NIS) measurements.\cite{NIS1}

The idea that both Coulomb interactions and the EP coupling are 
responsible for the VT was first explored by Ihle and Lorentz.\cite{IL1} 
They combined the tight-binding model introduced by Cullen and Callen 
\cite{CC1} with the Yamada model.\cite{yamada1} In their approach, the 
CO below $T_V$ is driven by the intersite electrostatic interactions 
and the condensation of the $\Delta_5$ phonon.\cite{IL2} Above $T_V$, 
the model explains the short-range polaronic order and provides a good 
description of the electric conductivity.\cite{IL3} Indeed, the 
polaronic character of charge carriers has been observed in the optical
conductivity\cite{Raman2,OC} and in photoemision studies.\cite{PES3} 
The cooperative mechanism of the VT was next studied also in the 
Peierls-Hubbard model,\cite{Seo} where strong on-site interaction 
induces the OO and lattice distortion. Since the active electrons on 
Fe$^{+2}$ ions occupy partly filled degenerate $t_{2g}$ states, the EP 
coupling could be further enhanced by the Jahn-Teller effect. This 
mechanism was confirmed by the dynamical mean-field theory (DMFT) 
combined with the DFT in the GGA+DMFT method,\cite{JT1} and by GGA+$U$ 
calculations,\cite{JT2} which showed that the local Jahn-Teller 
distortions modify the electronic structure lead indeed to opening 
of the insulating gap. 

In the previous paper,\cite{PRL} we reported the phonon spectrum obtained 
using the direct method,\cite{direct} and analyzed the EP interactions in 
presence of local Coulomb interaction $U$. We have found that two phonon 
modes with $X_3$ and $\Delta_5$ symmetry, which significantly distort the 
$B$-site octaedra, couple strongly to $t_{2g}$ electronic states. The $X_3$ 
phonon opens the gap at the Fermi energy, and drives the metal-insulator 
transition. Using the group theory, we proved that these two modes are 
primary order parameters describing the structural phase transition.

The purpose of the present paper is to focus on the details of the 
group theory analysis of the phase transformation in the magnetite, 
from the cubic $Fd\bar{3}m$ to monoclinic $P2/c$ symmetry, and to 
relate it to the simultaneous changes in the electronic structure and 
lattice dynamics. A full list of relevant OPs will be presented and 
their couplings will be discussed on the basis of Landau free energy. 
In the second part of the paper we focus on the results of structural 
optimization and lattice dynamics calculations in presence of local 
electron interactions, and demonstrate that the phase transformation in 
the VT involves the EP interaction.

The paper is organized as follows. In Sec. II the group theory analysis 
of the VT is presented. Sec. III is devoted to the electronic structure 
calculation, where we first present the method used (Sec. \ref{sec:gga}), 
and next analyze the optimized crystal structures (Sec. \ref{sec:opt}).
In Sec. IV we describe lattice dynamics and classify the obtained modes 
for the magnetite using the irreducible representations of the cubic 
group. These results serve as a basis to discuss the EP interactions and 
their effect on the electronic structure in Sec. V. We show a strong
coupling between the electronic properties and lattice dynamics, and in 
this context discuss the mechanism of the VT in Sec. VI.
The paper is concluded in Sec. VII, where we provide a coherent view on 
the VT in magnetite and summarize the presented results.

\section{Group theory analysis}

\subsection{Primary and secondary order parameters}
\label{sec:prima}

In a structural phase transition the symmetry is lowered from the 
high-symmetry phase, usually realized at high temperature, to the 
low-symmetry one, stable at low temperature. The Verwey transition in 
magnetite belongs to this class of phase transitions, so we consider 
first the Landau free energy functional to characterize the OPs 
responsible for the observed lowering of symmetry. In the Landau-type 
phase transition the space group ${\cal L}$ of the low-symmetry phase 
is a subgroup of the space group ${\cal H}$ of the high-symmetry phase, 
i.e., ${\cal L}\subset {\cal H}$. There exists often one irreducible 
representation (IR) of ${\cal H}$, which reduces the symmetry to 
${\cal L}$. This allows one to identify an OP as a vector in an IR 
space, $\Gamma_{{\bf k},j}(\eta_i)$, where ${\bf k}$ is a wave vector 
of the irreducible star, $j$ is the index of ray 
representation, and $\eta_i$ are the components of the OP.\cite{group}
The OP which defines the transition and determines the symmetry of the 
low-symmetry phase is called a {\it primary\/} OP, and the symmetry 
reduction diagram for it can be written as follows
\begin{equation}
{\cal H} \rightarrow [\Gamma_{{\bf k},j}(\eta_i)] \rightarrow {\cal L}.
\end{equation}

In contrast, a {\it secondary\/} OP $\gamma_{{\bf k},j}(\eta_i)$ is 
associated with an IR which reduces the symmetry of the high-symmetry 
space group to an intermediate space group ${\cal I}$, such that 
${\cal I}$ becomes a subgroup to ${\cal H}$, and a supergroup for 
${\cal L}$, i.e., ${\cal L}\subset {\cal I}\subset {\cal H}$.\cite{SOP}
Thus, the symmetry reduction diagram has the form
\begin{equation}
{\cal H} \rightarrow [\gamma_{{\bf k},j}(\eta_i)] \rightarrow {\cal I}.
\end{equation}
The secondary OP appears in the Landau free energy expansion always in 
a term which provides a linear coupling of this secondary OP to a 
quadratic (or higher power) expression in the relevant primary OP.
At the phase transition the primary OP starts to develop, while
the secondary OPs also become finite because of their coupling to the 
primary OP.

\begin{table}[b!]
\caption{
List of OPs from the parent space group $Fd3m$ (No=227) and basis 
$\{(1,0,0),(0,1,0),(0,0,1)\}$ to the monoclinic phase $P2/c$ (No=13, 
unique axis $b$, choice 1), basis $\{(\frac{1}{2},-\frac{1}{2},0), 
(\frac{1}{2},\frac{1}{2},0),(0,0,2)\}$ and origin 
$(\frac{1}{4},0,\frac{1}{4})$ relative to the original face center 
cubic lattice. Size is the ratio of the volumes of primitive 
low-symmetry to high-symmetry unit cells. The direction of the IR 
vector given in the last column indicates its nonvanishing components 
as well as the relations between the components of the OPs.}
\label{table1}
\begin{ruledtabular}
\begin{tabular}{ccccl}
IR & size & subgroup & No & IR vector\\
\hline
$\Gamma^+_1$ & 1 & $Fd\bar 3m$ & 227 & $(a)$            \cr
$\Gamma^+_3$ & 1 & $I4_1/amd$  & 141 & $(b,0)$          \cr
$\Gamma^+_4$ & 1 & $C2/m$      &  12 & $(c,c,0)$        \cr
$\Gamma^+_5$ & 1 & $Imma$      &  74 & $(d,0,0)$        \cr
$\Gamma^+_5$ & 1 & $C2/m$      &  12 & $(d,e,-e)$       \cr
$X_1$        & 2 & $Pmma$      &  51 & $(f,0,0,0,0,0)$  \cr
$X_3$        & 2 & $Pmna$      &  53 & $(g,0,0,0,0,0)$  \cr
$\Delta _2$  & 4 & $Pcca$      &  54 & $(0,0,0,0,h,-\frac{1}{\mu }h)$ \cr
$\Delta _4$  & 4 & $Pcca$      &  54 & $(0,0,0,0,p,\mu p)$            \cr
$\Delta _5$  & 4 & $Pbcm$      &  57 & $(0,0,0,0,0,0,0,0,q,-\mu q,-\mu q,-q)$ \cr
\end{tabular}
\end{ruledtabular}
\end{table}

\begin{table*}[t!]
\caption{
Subgroup relationships and the IRs creating the symmetry 
reduction from space group ${\cal H}$ to 
a given ${\cal I}$ and ${\cal L}$ space groups, respectively.
In brackets are the sizes of primitive unit cells.}
\label{table2}
\begin{ruledtabular}
\begin{tabular}{llcc}
Subgroup relationships & Group No's & Secondary OP & Primary OP \cr
\hline
$Fd\bar 3m(1)\;\supset\; I4_1amd(1)\;\supset\; Pmma(2)\;\supset\; Pbcm(4)$
    & 227, 141, 51, 57 & $\Gamma _3^+$, $X_1$ &  $\Delta _5$ \cr
$Fd\bar 3m(1)\;\supset\; Imma(1) \;\supset  Pmma(2) \;\supset\; Pcca(4)$
    & 227, 74, 51, 54  & $\Gamma _5^+$, $X_1$ &  $\Delta _2$,  $\Delta _4$ \cr
$Fd\bar 3m(1)\;\supset\; I4_1amd(1)\;\supset\; Pmna(2)$ 
    & 227, 141, 53     & $\Gamma _3^+$ & $X_3$ \cr
$Fd\bar 3m(1)\; \supset\; Imma(1)\; \supset\; Pmna(2)$
    & 227, 74, 53      & $\Gamma _5^+$ $(d,0,0)$ & $X_3$ \cr
$Fd\bar 3m(1)\;\supset\; Imma(1)\; \supset\;  C2/m(1)$
    & 227, 74, 12      & $\Gamma _5^+$ $(d,0,0)$ & $\Gamma _5^+$ $(d,e,-e)$ \cr
\end{tabular}
\end{ruledtabular}
\end{table*}

\begin{table*}[t!]
\caption{
Intersections of subgroups listed in Table I which lead to low-symmetry 
space group $P2/c$. In all cases the size is equal $4$, the basis vectors 
are $\{(\frac{1}{2},-\frac{1}{2},0),(\frac{1}{2},\frac{1}{2},0),
(0,0,2)\}$ and the origin is $(\frac{1}{4},0,\frac{1}{4})$.}
\label{table3}
\begin{ruledtabular}
\begin{tabular}{rlll}
No &IRs & subgroups giving $P2/c$ & subgroup No's \cr
\hline
 1 &$X_3$,$\Delta_5$ & $Pmna\;\cap\; Pbcm$  &  (53, 57) $=$ 13\cr
 2 &$X_3$,$\Delta_2$ & $Pmna\;\cap\; Pcca$  &  (53, 54) $=$ 13\cr
 3 &$X_3$,$\Delta_4$ & $Pmna\;\cap\; Pcca$  &  (53, 54) $=$ 13\cr
 4 &$X_3$,$\Delta_5$,$\Gamma _5^+$ & $Pmna\;\cap\; Pbcm\;\cap\; C2/m $ & (53, 57, 12)$=$ 13\cr
 5 &$X_3$,$\Delta_2$,$\Gamma _5^+$ & $Pmna\;\cap\; Pcca\;\cap\; C2/m $ & (53, 54, 12)$=$ 13\cr
 6 &$X_3$,$\Delta_4$,$\Gamma _5^+$ & $Pmna\;\cap\; Pcca\;\cap\; C2/m $ & (53, 54, 12)$=$ 13\cr
 7 &$X_3$,$\Delta_5$,$\Delta _2$   & $Pmna\;\cap\; Pbcm\;\cap\; Pcca $ & (53, 57, 54)$=$ 13\cr
 8 &$X_3$,$\Delta_5$,$\Delta _4$ & $Pmna\;\cap\; Pbcm \;\cap\;  Pcca $ & (53, 57, 54)$=$ 13\cr
 9 &$X_3$,$\Delta_2$,$\Delta _4$ & $Pmna\;\cap\; Pcca \;\cap\;  Pcca $ & (53, 54, 54)$=$ 13\cr
10 &$X_3$,$\Delta_5$,$\Delta _2$,$\Gamma _5^+$ & $Pmna\;\cap\; Pbcm\;\cap\; Pcca\;\cap\; C2/m$ 
   &  (53, 57, 54, 12) $=$ 13\cr
11 &$X_3$,$\Delta_5$,$\Delta _4$,$\Gamma _5^+$ & $Pmna\;\cap\; Pbcm\;\cap\; Pcca\;\cap\; C2/m$ 
   &  (53, 57, 54, 12) $=$ 13\cr
12 &$X_3$,$\Delta_2$,$\Delta _4$,$\Gamma _5^+$ & $Pmna\;\cap\; Pcca\;\cap\; Pcca\;\cap\; C2/m$
   &  (53, 54, 54, 12) $=$ 13\cr
13 &$X_3$,$\Delta_5$,$\Delta _2$,$\Delta _4$   & $Pmna\;\cap\; Pbcm\;\cap\; Pcca\;\cap\; Pcca$
   &  (53, 57, 54, 54) $=$ 13\cr
14 &$X_3$,$\Delta_5$,$\Delta _2$,$\Delta _4$,$\Gamma _5^+$ 
   & $Pmna\;\cap\; Pbcm\;\cap\;Pcca\;\cap Pcca\;\cap C2/m $ & (53, 57, 54, 54, 12)$=$ 13\cr
\end{tabular}
\end{ruledtabular}
\end{table*}

In principle a phase transition could be induced by two or more 
primary OPs, and they can be coupled to the secondary ones. Thus, 
in general $N$ primary OPs, IR$\,_1$,IR$\,_2$,$\dots,$IR$\,_N$,
reduce the high-symmetry space group ${\cal H}$ into ${\cal L}_1$, 
${\cal L}_2$, $\dots,$ ${\cal L}_N$ low-symmetry space groups, 
respectively,
\begin{eqnarray}
&{\cal H}&\rightarrow [\Gamma_{{\bf k}_1,j_1}(\eta_i)]\rightarrow {\cal L}_1, \nonumber\\
&{\cal H}&\rightarrow [\Gamma_{{\bf k}_2,j_2}(\eta_i)]\rightarrow {\cal L}_2, \nonumber\\
&\cdots&                                                                      \nonumber\\
&{\cal H}&\rightarrow [\Gamma_{{\bf k}_N,j_N}(\eta_i)]\rightarrow {\cal L}_N.
\end{eqnarray}
Due to a non-linear coupling between the primary OPs, the final 
low-symmetry space group ${\cal L}$ is an intersection of all low-symmetry
space groups
\begin{equation}
{\cal L} = {\cal L}_1 \cap {\cal L}_2 \cap \dots {\cal L}_N.
\end{equation}
Thus, the low-symmetry space group ${\cal L}$ consists of those 
symmetry elements which are present in {\it all subgroups\/} 
${\cal L}_1,{\cal L}_2, \dots, {\cal L}_N$.
The secondary OPs are associated with IRs which reduce the symmetry of 
the high-symmetry space group to intermediate space groups ${\cal I}_m$, 
such that ${\cal L}_n\subset {\cal I}_m\subset {\cal H}$, where 
$n=1,2,\dots N$ and $m$ should be found by inspection the group-subgroup 
relationships
\begin{eqnarray}
&{\cal H}&\rightarrow [\gamma_{{\bf k}_1,j_1}(\eta_i)]
          \rightarrow {\cal I}_1, \nonumber\\
&{\cal H}&\rightarrow [\gamma_{{\bf k}_2,j_2}(\eta_i)]
          \rightarrow {\cal I}_2, \nonumber\\
&\cdots&                          \nonumber\\
&{\cal H}&\rightarrow [\gamma_{{\bf k}_m,j_m}(\eta_i)]
          \rightarrow {\cal I}_m.
\end{eqnarray}
For instance, a symmetry reduction in shape memory alloy NiTi, 
associated with the symmetry change $Pm{\bar 3}m\rightarrow P2_1/m$, 
is a phase transition which involves two primary OPs.\cite{hatch,magda} 
As we demonstrate in this Section, the symmetry reduction realized in 
the VT of magnetite requires also multiple primary and secondary OPs.

The high-symmetry space group of Fe$_3$O$_4$ above $T_V$ is
${\cal H}=Fd{\bar 3}m$ (Fig. 1). The low-symmetry phase ${\cal L}$ was 
identified as the monoclinic $Cc$ symmetry 
\cite{struct2,struct3,struct4,struct5} with a unit cell of approximate 
size $\sqrt2 a\times\sqrt2 a\times 2a$. As argued by Iizumi {\it et al.},
\cite{struct3} a smaller cell $a/\sqrt2 \times a/\sqrt2 \times 2a$ with 
the $Pmca$ symmetry can be used for refinement almost all Bragg 
reflections. In Ref. \onlinecite{attfield2}, the monoclinic symmetry 
$P2/c$ with orthorhombic constraints was used in the refinement procedure. 
This symmetry describes all Bragg peaks except for three very weak 
reflections arising from atomic displacements smaller than $0.01$ \AA{}.
\cite{attfield2} Indeed, the unit cell used in the refinement procedure 
takes into account all atomic displacements coming from 
$\textbf{k}_{\Gamma}$, $\textbf{k}_{\Delta}$, and $\textbf{k}_X$ points. 
Therefore, the $P2/c$ symmetry explains majority of physical properties 
connected with the VT. 

Here, we present a detailed group theory analysis of the structural 
transition for ${\cal L}=P2/c$ (No=13), with basis $\{(\frac{1}{2},
-\frac{1}{2},0),(\frac{1}{2},\frac{1}{2},0),(0,0,2)\}$ and the origin 
$(\frac{1}{4},0,\frac{1}{4})$ relative to the original face center 
cubic lattice. Using the {\sc copl} and {\sc isotropy} computer codes 
\cite{copl,isotropy} we obtained the lists of possible intermediate 
$\{{\cal I}_m\}$ and low-symmetry $\{{\cal L}_n\}$ space groups, and 
show them in Table I. The space groups, which have been found, are 
reduced by IRs from the high-symmetry space group ${\cal H}=Fd{\bar 3}m$.
As is clearly seen, there is no IR which reduces $Fd{\bar 3}m$ to 
low-symmetry ${\cal L}=P2/c$.
This means that the phase transition in Fe$_3$O$_4$ is driven by 
{\it at least\/} two primary OPs. A closer inspection of space 
group-subgroup relationships, partly shown in Table II, leads us to 
the conclusion that in principle there are five primary OPs: 
$X_3$, $\Delta_2$, $\Delta_4$, and $\Delta_5$, and $\Gamma_5^+$ 
($T_{2g}$) within direction $(d,e,-e)$. 
Indeed, the intersection of space groups being the result of symmetry 
reduction of $Fd{\bar 3}m$ by the above mentioned IRs provides 
the observed low-symmetry space group $P2/c$, Table III.

In the the simplest case, the $P2/c$ symmetry can be induced by two OPs.
Taking the $X_3$ and $\Delta_5$ symmetries, the reduction diagrams can
be written
\begin{eqnarray}
&Fd{\bar3}m&\rightarrow [X_3, {\bf k}=(0,0,1)] \rightarrow Pmna,  \nonumber\\
&Fd{\bar3}m&\rightarrow \left[\Delta_5, 
            {\bf k}=\left(0,0,\frac{1}{2}\right)\right]\rightarrow Pbcm. 
\end{eqnarray}
Now, if we take common symmetry elements (an intersection) of these 
two groups, we get
\begin{equation}
Pmna \cap Pbcm = P2/c,
\end{equation}
so the low-symmetry space group has indeed $P2/c$ symmetry.
The remaining IRs involved in the phase transition $\Gamma^+_1$ ($A_{1g}$), 
$\Gamma^+_3$ ($E_g$), $\Gamma^+_4$ ($T_{1g}$), $\Gamma _5^+$ ($T_{2g}$) 
within direction $(d,0,0)$, and $X_1$ are the secondary OPs. On the one 
hand, it is evident that $\Gamma_5^+$ can be either primary or secondary
OP. If only one component $d$ is nonzero it has a secondary character 
and its reduction diagram reads
\begin{equation}
Fd{\bar3}m \rightarrow [\Gamma_5^+(d,0,0),{\bf k}=(0,0,0)]\rightarrow Imma.
\end{equation}
The resulting group fulfills the following subgroup relationship
\begin{equation}
Fd{\bar 3}m \supset Imma \supset  Pmna,
\end{equation}
so $\Gamma_5^+$ as the secondary OP couples to the primary OP $X_3$.
On the other hand, if all three components of $\Gamma_5^+$ are nonzero,
it has primary character with the reduction diagram
\begin{equation}
Fd{\bar3}m\rightarrow [\Gamma_5^+(d,e,-e),{\bf k}=(0,0,0)]\rightarrow C2/m.
\end{equation}
Because of the following relationship
\begin{equation}
Fd{\bar 3}m \supset Imma \supset  C2/m,
\end{equation}
$\Gamma_5^+(d,0,0)$ is the secondary parameter of the $\Gamma_5^+(d,e,-e)$ OP.
In what follows, we shall use the point group notation of the IRs at the 
$\Gamma$ point.
This notation is usually used in experimental works on phonon spectroscopy.

As is clear from this analysis, the $X_3$ OP plays an exceptional role among 
other symmetries, being present in all intersections listed in Table III. 
So one recognizes that this OP is essential to generate the LT
monoclinic phase and it confirms its precursor behavior found in the diffuse 
scattering study. As we showed by the {\it ab initio\/} 
calculations, this mode strongly couples to electrons 
and induces the metal-insulator transition. \cite{PRL}

At the $\textbf{k}_{\Delta}$ point, there are three possible OPs: 
$\Delta_2$, $\Delta_4$, and $\Delta_5$. In principle, by coupling to 
$X_3$ either of these modes can induce $P2/c$ symmetry, but one of them, 
$\Delta_5$, has a strong support from the experiment. Originally proposed 
by Yamada\cite{yamada1} --- this mode was found to be the main component 
of lattice distortion in critical neutron scattering.\cite{diff1} Neutron 
studies by Iizumi {\it et al.}\cite{struct3} confirmed the $\Delta_5$ 
pattern of displacements, which gives rise to $\textbf{k}_{\Delta}$ Bragg 
peaks. The EP interaction for the $\Delta_5$ mode was studied within the 
model of Ihle and Lorenz.\cite{IL1} At the $\Gamma$ point there could be 
only one primary OP with $T_{2g}$ symmetry.
It is related with the softening of the elastic constant $c_{44}$.
Notice that in the cubic system one has $c_{44}=c_{55}=c_{66}$, and 
therefore one is unable to differentiate between the secondary $(d,0,0)$ 
or primary $(d,e,-e)$ OP characters. A strong EP interaction for the 
$T_{2g}$ mode was found by the Raman experiment \cite{Raman3}
and in neutron diffuse measurements.\cite{yamada2} Taking into account 
that only $t_{2g}$ electronic states are present near the Fermi level, 
a strong coupling to phonons with the same symmetry is not surprising. 
Altogether, from experimental observations and from {\it ab initio\/} 
calculations, presented in detail in the next Sections, we have strong 
evidence for the following three primary OPs being involved in the phase 
transition: $X_3$, $\Delta _5$, and $T_{2g}$. 

To appreciate fully the complexity of the resulting LT phase, 
we emphasize that although the LT symmetry is uniquely 
determined by the primary OPs, the exact atomic positions result from 
all lattice distortions, associated also with secondary OPs. The $X_1$ 
type distortion was reported in x-ray\cite{X1} and in neutron
\cite{struct3} diffraction. The mechanism stabilizing the incommensurate 
mode $\Delta_5$ at wavevector $\textbf{k}_{\Delta}$ by simultanuous 
condensation with the $X_1$ phonon was discussed by Iizumi.\cite{X1D5}
The $X_1$ mode induces the longitudinal displacements with $\lambda=a$ 
observed in the monoclinic structure.\cite{attfield2}
There are a few secondary OPs at the zone center, and they are likely 
to participate in the LT structure as well.\cite{struct3,attfield2}
For example, the anomaly related to the observed changes in frequency 
and linewidth at the phase transition was found for the highest Raman 
$A_{1g}$ mode.\cite{Raman2}

\subsection{Free energy functional}
\label{sec:freen}

Knowing the primary and secondary OPs and their IRs, one can construct the
invariant quantities which do not change under symmetry group transformations,
and next expand the free energy ${\cal F}$ into a series of the 
components which involve different OPs.\cite{isotropy} Each component is 
a linear combination of its electronic and phononic parts, which represent 
the electronic density and atomic displacement contributions, respectively. 
For instance, the component of $X_3$ can be written in the form which 
includes its electronic and phononic contribution: 
$g=g_{\rm el}+g_{\rm ph}$. Note that in the present case of active $t_{2g}$ 
degrees of freedom of $3d$ electrons at iron ions, the electronic part 
$g_{\rm el}$ may as well involve the orbital occupations. 

We limit the expansion of the free energy ${\cal F}$ to the 4-th order
terms, with the exception of two primary OPs $X_3$ and $\Delta _5$, for 
which we include also terms of 6-th order as they are typical for the 
discontinuous (first-order) phase transitions. A complete expansion gives 
too lengthy expresion to be reproduced here, so we leave in the expansion
only components of OPs required by symmetry as shown in Table I, and write 
down only the potentially nonvanishing terms. Each term of the series 
must be multiplied by factors depending on thermodynamical parameters, 
however, we shall not discuss this aspect here. The free energy reads

\begin{widetext}

\begin{eqnarray}
  \label{free}
{\cal F} &=& (g^2 + q^2 + g^4 + q^4 + g^2q^2 + g^6 + g^4q^2 + g^2q^4 + q^6)
 + (p^2 + h^2 + g^2p^2 + g^2h^2 + q^2p^2 + q^2h^2 + p^2h^2 + gq(h + p)) 
\nonumber \\ 
&+&(f^2 + fq^2 + fp^2+ fh^2 + fph + f^2g^2) + (a^2 + ag^2 + aq^2 + ap^2 + ah^2) 
\nonumber \\
&+&(b^2 + b^3 + bg^2 + bq^2 + bp^2 + bh^2) 
 + (c^2 + c^2g^2 + c^2q^2) + (d^2 + e^2 + dg^2 + dp^2 + dh^2 + dph + ep^2 + eh^2).
\end{eqnarray}

\end{widetext}

First bracket in Eq. (\ref{free}) corresponds to the $X_3$ and $\Delta _5$ 
IRs and their couplings (see Table I). We anticipate that the lowest 
coupling term between them $g^2q^2$ must be large and should lower the 
free energy ${\cal F}$. Second bracket correponds to $\Delta_2$ and 
$\Delta_4$ IRs and to the coupling terms among them. The second line in Eq. 
(\ref{free}) describes the coupling of IR $X_1$ with the primary OPs, 
being of linear-quadratic form. The interaction between the $X_1$ and 
$\Delta_5$ phonons discussed by Iizumi \cite{X1D5} is described by the 
$fq^2$ term. The following terms given in fourth, fiveth, sixth and 
seventh brackets determine the coupling of IRs of the crystal point groups 
$A_{1g}$, $E_g$, $T_{1g}$, and $T_{2g}$, respectively, with the primary 
OPs. Again their forms are adequate to describe the influence of 
secondary OPs on the transition. In particular, the term $dg^2$ in the 
last line, describing the coupling between the $X_3$ and $T_{2g}$ OPs,
contributes to the softening of the $c_{44}$ elastic constant. 
\cite{elastic2} Other couplings involving the zone-center IRs may 
account for anomalies observed by the Raman studies.\cite{Raman2,Raman3}

In order to identify the relevant part of the free energy Eq. 
(\ref{free}) which drives the VT in magnetite, we consider next 
the changes of the electronic structure (Sec. III) and the phonon 
spectra (Sec. IV) at the transition. In fact, we shall demonstrate 
below that these properties are strongly coupled to each other, and 
the VT is driven by the electron-phonon interaction enhanced by the 
local Coulomb interaction $U$, see Sec. V.

\section{Electronic structure}

\subsection{Computational methods}
\label{sec:gga}

The lattice parameters of Fe$_3$O$_4$ were optimized together with the
electronic structure using the total energy DFT approach. The numerical 
calculations were performed by means of the {\sc vasp} program
\cite{vasp} within the GGA+$U$ approach.\cite{gga} The program uses 
highly accurate full-potential projector-augmented wave 
(PAW) method, originally proposed by Bl\"{o}chl \cite{paw} and 
implemented by Kresse and Joubert.\cite{pawvasp} The wavefunctions in 
the core region are obtained by the linear transformation from 
variational pseudo-wavefunctions. In the augmented region the 
wavefunctions are expanded in the plane wave basis. Two supercells with
the periodic boundary conditions were chosen: $a\times a\times a$ and 
$a/\sqrt2\times a/\sqrt2\times 2a$ for the $Fd\bar{3}m$ and $P2/c$ 
structures, respectively, with 56 atoms each. The latter supercell was 
used also for the reference system of $Fd\bar{3}m$ symmetry, when total
energies of two considered structures were compared.

For a given ionic configuration, the Kohn-Sham Hamiltonian was
diagonalized by the iterative residual minimization method,\cite{vasp}
which allows one for efficient optimization of the charge density and
wavefunctions. In the initial steps also the blocked Davidson minimization
method was applied. To improve convergence a small smearing of electronic 
states at the Fermi surface was introduced with the effective parameter
$\sigma=0.2$ eV. The Monkhorst-Pack scheme \cite{MP} was applied 
for summation over two ${\bf k}$-point grids $6\times 6\times 6$ and 
$6\times 6\times 2$ relevant for $Fd\bar{3}m$ and $P2/c$ symmetries, 
respectively, which are large enough for a meaningful energy comparison. 
In Ref. \onlinecite{PRL} we used the smaller grids, 
i.e., $4\times 4\times 4$ and $4\times 4\times 2$, 
which give already satisfactory values for lattice parameters and energies. 
However, larger ${\bf k}$-point grids were used at present to reach a 
higher accuracy as they give even more reliable values of force constants 
and phonon frequencies. The energy cut-off for the plane wave expansion 
was set at $520$ eV after we have verified that the contribution of 
states with higher energies was negligible.

The local electron interations between $3d$ electrons on Fe ions were 
included in the Hartree-Fock approximation, as usually done in the LDA+$U$
method.\cite{LDAU1} As a result one finds the total energy of 
the system $E_{\rm tot}$ in the form
\begin{equation}
\label{etot}
E_{\rm tot}=E_{\rm GGA}+E_U-E_{\rm dc},
\end{equation}
where $E_{\rm GGA}$ is the energy obtained in the GGA approach, $E_U$ 
describes the contribution due to local interactions parametrized by 
the Coulomb element $U$ and Hund's exchange $J$, and $E_{\rm dc}$ is 
the double counting correction term, i.e., the averaged 
electron-electron interaction, which has to be subtracted. The details 
of the Hamiltonian used in the calculations were presented in Ref. 
\onlinecite{FeU}. 

In the present calculations we have chosen the following parameters: 
$U=4.0$ eV and $J=0.8$ eV. The value of $U$ agrees with the constrained 
DFT calculations,\cite{DFT2,JT1} and 
is somewhat reduced from the ionic value of 6.4 eV estimated by Zaanen 
and Sawatzky.\cite{Zaa90} The value of $J$ was obtained from the atomic 
values of the Racah parameters for Fe$^{2+}$ ions:\cite{Zaa90} $B=0.131$ 
eV and $C=0.484$ eV. Here we use an average value of Hund's exchange,
\cite{Ole05} $J=\frac{5}{2}B+C$, as usually implemented within the LDA+$U$ 
method.\cite{Lie95} Note that the value of $J=1.0$ eV used in some LSDA+$U$
calculations\cite{Leonov} and in the LDA+DMFT method\cite{JT1} is somewhat 
overestimated (in fact this value applies to a pair of $e_g$ electrons 
rather than to a pair of $t_{2g}$ electrons). The nearest-neighbor 
Coulomb interaction is one order of magnitude smaller $V\simeq 0.3-0.4$
eV,\cite{DFT2} and we assume that it is to a large extent included 
already in the GGA scheme. 

The optimal atomic configuration, which gives the minimum of the total 
energy, $E_0=min\{E_{\rm tot}\}$, was found using the conjugate gradient 
and quasi-Newton procedures.\cite{vasp} Because of phonon calculations, 
which are based on the well-optimized supercell, the terminating criteria 
for the electronic and ionic degrees of freedom were very strict: 
$10^{-7}$ eV and $10^{-5}$ eV, respectively. To get more precise
atomic positions we used also force criterion, in quasi-Newton procedure, 
which allowed us to get residual forces less than $10^{-2}$ meV/\AA{}
(the pressure was then less than $0.1$ kbar).

Phonon frequencies were calculated only for the cubic structure, using 
the direct method \cite{direct} implemented in the {\sc phonon} program. 
\cite{phonon} According to the Hellmann-Feynmann (HF) theorem, the 
atomic forces can be obtained by displacing atoms from their equilibrium 
positions. The minimum number of displacements depends on crystal 
symmetry and on the number of nonequivalent atoms. For magnetite only 
three independent displacements for Fe($A$), Fe($B$), and O atoms are 
sufficient. To minimize systematic errors we performed displacements 
in positive and negative directions. 

The combination of derived displacements and forces allows 
one to obtain the force constants matrix elements by the singular value 
decomposition method. A special treatment was applied for calculations 
with finite $U$, where the convergence of the electronic part was in some 
cases very slow. When this occured, the calculations of the HF forces were 
initialized using the wavefunctions optimized for the cubic structure. 
For magnetite, the direct method provides exact frequencies at the 
$\Gamma$ and $X$ points. The phonon frequencies at other points were 
evaluated with only small errors due to large supercell used in the 
present calculations.

\subsection{Crystal optimization}
\label{sec:opt}

The structural properties of the cubic and monoclinic phases have been 
discussed in many previous studies. In Tabs. IV and V we summarize the 
electronic and lattice parameters obtained in the present calculations, 
comparing them with the experimental data. We focused mainly on the 
effects connected with the Coulomb interaction $U$. For the cubic symmetry, 
the lattice constant $a$ and the internal oxygen parameter $x$ depend 
rather weakly on $U$, and show overall good agreement with experimental
data of Refs. \onlinecite{attfield2,OC,momentA,moment_tot} (Table IV). 
As expected and noticed before,\cite{LDAU3} the magnetic moments on 
particular atoms increase with $U$, improving the agreement between the 
electronic structure calculations and the experimental data. The average 
magnetic moment per one formula unit $m_{\rm av}$, however, does not 
change with $U$ and $J$. It shows that magnetic polarization in this 
material is already at maximum and can be therefore well described in 
the band picture, while reliable calculations concerning the magnetic
polarization of individual Fe ions require local electron interactions
$\{U,J\}$. One finds that magnetite is metallic ($\Delta_g=0$) in the 
cubic phase, independently of the actual values of $U$ and $J$. Note 
that the total ground state energy of the crystal $E_{\rm tot}$ 
increases with $U$, but direct comparison between the energies 
obtained in the GGA and the GGA+$U$ is meaningless. 

\begin{table}[t!]
\label{gs}
\caption{
Ground state lattice and electronic parameters for $Fd\bar{3}m$ and 
$P2/c$ structures. The magnetic moments of individual ions Fe($A$) 
($m_A$), Fe($B$) ($m_B$), and O ($m_O$) obtained within the present 
GGA ($U=J=0$) and GGA+$U$ ( with $U=4.0$ eV and $J=0.8$ eV) calculations 
for the high-temperature $Fd3$m phase are compared with the experimental 
data, wherever available. The lattice constants ($a,b,c$) are in \AA, 
magnetic moments $\mu_i$ in $\mu_B$, and energies $E_{\rm tot}$ and 
the gap $\Delta_g$ in eV.}
\begin{ruledtabular}
\begin{tabular}{c c c c c}
phase & quantity & GGA & GGA+$U$ & experiment  \cr
\hline
$Fd\bar{3}m$ & $a$     & 8.377    & 8.446   & $8.394^a$   \cr
       & $x$           & 0.2545   & 0.2549  & $0.2549^a$  \cr
       & $m_A$         &$-$3.40   &$-$3.97  & $-3.82^b$   \cr
       & $m_B$         &  3.52    & 3.87    &             \cr
       & $m_O$         & 0.08     & 0.04    &             \cr
       & $m_{\rm av}$  & 3.96     & 3.96    & $4.05^c$    \cr
       & $E_{\rm tot}$ &$-$426.39 &$-$386.80&             \cr
       & $\Delta_g$    & 0        &  0      &             \cr
\hline
$P2/c$ & $a$ & 5.926   & 5.967    & $5.944^a$             \cr
       & $b$ & 5.926   & 5.995    & $5.925^a$             \cr
       & $c$ & 16.752  & 17.034   & $16.775^a$            \cr
       & $\beta$       & 90.004   & 90.417   & $90.237^a$ \cr
       & $E_{\rm tot}$ &$-$426.63 &$-$389.28 &            \cr
       & $\Delta_g$    & 0        &  0.33    &  $0.14^d$  \cr
\end{tabular}
\end{ruledtabular}
\leftline{$^a$ Reference \onlinecite{attfield2}} 
\leftline{$^b$ Reference \onlinecite{momentA}} 
\leftline{$^c$ Reference \onlinecite{moment_tot}}
\leftline{$^d$ Reference \onlinecite{OC}} 
\end{table}

The $P2/c$ structure was optimized starting from the experimentally 
determined geometry,\cite{attfield2} but we did not impose orthorhombic 
$Pmca$ symmetry constraints on atomic positions. The theoretical crystal 
structures, obtained also by the {\sc vasp} program, were already 
discussed in detail in Ref. \onlinecite{Jeng}. We note that some 
discrepancies between the results of these two calculations may follow 
from different energy cut-off and termination conditions, as well as from 
slightly different values of $U$ and $J$ used. In the GGA calculations 
performed with $U=J=0$, the lattice constants, atomic positions, and the 
total energy converge to those of the high-symmetry cubic structure 
(Tabs. IV and V). This behavior agrees with the molecular dynamics 
simulations presented in Ref. \onlinecite{JT2}. 
The proximity of the total energies $E_{\rm tot}$ obtained in both 
phases indicates that by going beyond the GGA one of these structures 
will be stabilized. Note that the energy difference between these two 
phases obtained in the GGA calculations is too small to arrive at 
definite conclusion concerning the stability of the LT phase of $P2/c$ 
symmetry.

\begin{table*}[t!]
\label{geo}
\caption{Atomic positions for iron ions (at $A$ and $B$ sites) and oxygen 
ions in the $P2/c$ structure, as obtained in the GGA calculations (left, 
$U=J=0$) and in the GGA+$U$ calculations (middle, with $U=4.0$ eV and 
$J=0.8$ eV), compared with the experimental data of Ref. 
\onlinecite{attfield2} (right).
}
\begin{ruledtabular}
\begin{tabular}{c d d d d d d d d d}
& \multicolumn{3}{c}{$U=J=0$} &   \multicolumn{3}{c}{$U=4.0$ eV, $J=0.8$ eV} & \multicolumn{3}{c}{Experiment\cite{attfield2}}\cr
\hline  
atom  & $x$ & $y$ &  $z$ & $x$ & $y$ & $z$ & $x$ & $y$ & $z$ \cr
\hline
$A1$  &  0.2490 & 0.0 & 0.0625 & 0.2491  & 0.0104 & 0.0637 & 0.25 & 0.0034 & 0.0637 \cr
$A2$  &  0.25   & 0.5 & 0.1875 & 0.2493  & 0.4991 & 0.1885 & 0.25 & 0.5061 & 0.1887 \cr
$B1a$ &  0.0    & 0.5 & 0.0    & 0.0     & 0.5    & 0.0    & 0.0  & 0.5    & 0.0    \cr
$B1b$ &  0.5    & 0.5 & 0.0    & 0.5     & 0.5    & 0.0    & 0.5  & 0.5    & 0.0    \cr
$B2a$ &  0.0    & 0.0 & 0.25   & 0.0     & 0.0050 & 0.25   & 0.0  & 0.0096 & 0.25   \cr
$B2b$ &  0.5    & 0.0 & 0.25   & 0.5     & 0.9996 & 0.25   & 0.5  & 0.0096 & 0.25   \cr
$B3$  &  0.25   & 0.25 & 0.375 & 0.2502  & 0.2604 & 0.3804 & 0.25 & 0.2659 & 0.3801 \cr
$B4$  &  0.25   & 0.75 & 0.375 & 0.2529  & 0.7553 & 0.3740 & 0.25 & 0.7520 & 0.3766 \cr
\hline
O1  &  0.25  & 0.2591& -0.0023& 0.2508 &0.2689& 0.0000 & 0.25    &0.2637& -0.0023 \cr
O2  &  0.25  & 0.7408& -0.0023& 0.2491 &0.7572&-0.0026 & 0.25    &0.7461& -0.0029 \cr
O3  &  0.25  & 0.2409& 0.2523 & 0.2487 &0.2370& 0.2541 & 0.25    &0.2447& 0.2542  \cr
O4  &  0.25  & 0.7591& 0.2523 & 0.2482 &0.7708& 0.2474 & 0.25    &0.7738& 0.2525  \cr
O5a &-0.0091 & 0.0   & 0.1272 &-0.0155 &0.0045& 0.1297 & -0.0091 &0.0095& 0.1277  \cr
O5b & 0.4909 & 0.0   & 0.3727 & 0.4941 &0.0187& 0.3701 & 0.4909  &0.0095& 0.3723  \cr
O6a &-0.0092 & 0.5   & 0.1227 &-0.0083 &0.4911& 0.1257 & -0.0081 &0.5046& 0.1246  \cr
O6b & 0.4908 & 0.5   & 0.3772 & 0.4842 &0.5042& 0.3730 & 0.4919  &0.5046& 0.3754  \cr
\end{tabular}
\end{ruledtabular}
\end{table*}

Indeed, the optimization performed with $U=4.0$ eV and $J=0.8$ eV gives 
lower total energy of the $P2/c$ structure by about $2.48$ eV compared
to the cubic phase. This result confirms that the monoclinic structure 
is more stable in the low temperature regime. The obtained insulating 
gap $\Delta_g=0.33$ eV is larger than the experimental value $0.14$ eV,
but is close to the values obtained in other electronic structure 
calculations.\cite{JT2}
One should note that the calculated gap is usually larger for the
relaxed system than for the experimental geometry. 

The charge and spin distribution in the $3d$ states on Fe($B$) ions is 
presented in Table VI. Since we have chosen larger Wigner-Seitz radius 
$r_{\rm WS}=1.5$ \AA{}, the obtained values are somewhat larger than in other 
studies.\cite{prl1,prl2} However, the difference between the larger 
($B1$ and $B4$) and smaller ($B2$ and $B3$) charges ($\sim 0.2e$) is similar, 
and agrees well with experiment.\cite{attfield2} The electron density 
of states (DOS) and the orbital density distribution will be analyzed 
in Sec. V, where the EP interactions are discussed.

The parameters of the relaxed crystal structure show good agreement with 
experiment.\cite{attfield2} The lattice constants are overestimated only 
by about $1\%$, and the monoclinic angle is well reproduced. As in the 
real crystal of magnetite, the atoms are displaced predominantly along 
the $y$ direction with respect to the original cubic positions. In the 
monoclinic cell, the $y$ direction corresponds to the [110] diagonal one 
in the $Fd\bar{3}m$ symmetry. According to the experiment,\cite{attfield2} 
the largest atomic shift of $0.13$ \AA{} is found for Fe($B3$) ions, and 
it agrees well with the theoretical value $0.11$~\AA{}. Displacements 
of Fe($B2$) and Fe($B4$) ions are much smaller ($0.03$-$0.04$~\AA), 
and Fe($B1$) ions do not change their positions at all. As observed 
experimentally, the Fe($A$) ions have only minor displacements 
($\sim0.03$ \AA), and they are smaller than those found theoretically 
($0.05$ \AA{}). The largest shift of oxygen ions amounts to $0.11$~\AA{}, 
and is slightly larger than the experimental value found for O4 ions, 
being $0.09$~\AA{}. 

This present analysis shows that the largest atomic displacements, 
leading to the monoclinic phase, are connected with Fe($B$) and O ions.
They induce deformations of Fe$_B$O$_6$ octahedra. It shows that models 
which assume only oxygen vibrations are probably not sufficient to 
describe fully all relevant degrees of freedom involved in the Verwey 
phase transition. In Secs. IV and V we will study the phonon spectrum, 
and in particular we will focus on those modes, which strongly 
influence the electronic and crystal structure. 

\begin{table}[b!]
\label{tab:j}
\caption{
Gap in the electronic structure $\Delta_g$ (in eV), and the charge and 
magnetization densities for nonequivalent Fe($B$) atoms, $n_{Bi}$ 
and $m_{Bi}$, as obtained in the $P2/c$ phase with $U=4.0$ eV for 
two values of Hund's exchange: $J=0.8$ eV (left) and $J=0$ (right). }
\begin{ruledtabular}
\begin{tabular}{c c c c c c c}
&   $J$      &  \multicolumn{2}{c}{0.8}  &  \multicolumn{2}{c}{0.0}  \cr
& $\Delta_g$ & \multicolumn{2}{c}{0.33}  & \multicolumn{2}{c}{0.54}  \cr
\hline
&  atom $B$  &$n_{Bi}$&$m_{Bi}$&$n_{Bi}$&$m_{Bi}$\cr
\hline
&     $B1$   &  6.07  &  3.64  &  6.09  &  3.66  \cr
&     $B2$   &  5.85  &  4.05  &  5.82  &  4.15  \cr
&     $B3$   &  5.87  &  4.00  &  5.85  &  4.09  \cr
&     $B4$   &  6.09  &  3.62  &  6.11  &  3.64  \cr
\end{tabular}
\end{ruledtabular}
\end{table}

\subsection{Effect of Hund's exchange}
\label{sec:J}

We close this Section by investigating the effect of Hund's exchange on
the electronic structure for fixed $U=4.0$ eV. One finds that the gap 
value is considerably enhanced for $J=0$ ($\Delta_g=0.54$ eV) over the 
value found for the realistic exchange interaction $J=0.8$ eV 
($\Delta_g=0.54$ eV), see Table VI. This demonstrates that the gap 
opens between the minority subbands and the actual effective interation 
between the electrons with the same spins is reduced by the exchange 
term. When Hund's exchange is missing, this interaction is enhanced, 
resulting in an increased gap value. Note that the effective 
interaction $(U-J)$ between the electrons of the same spin decides also 
about the magnetic and phonon properties of the iron metal.\cite{FeU}

The charge density distribution $\{n_{Bi}\}$ over the iron ions in $B$ 
positions is only little modified when $J=0$ is selected. In fact, the 
polarization between the amplitude of the CO between the $\{B1,B4\}$
and $\{B2,B3\}$ pairs of ions is somewhat increased which again
confirms that the effective interaction between the minority 
electrons has increased. Consequently, the magnetic moments $m_{Bi}$ 
are also slightly increased for $J=0$ over their values found for
$J=0.8$ eV. This behavior allows us to conclude that the calculations 
\cite{JT1,Leonov} perfomed with the same value of $U=4.0$ eV but with 
larger $J=1.0$ eV correspond in fact to {\it weaker\/} 
interaction between the minority electrons.

\section{Lattice dynamics}

\begin{table}[b!]
\label{tab:phonons}
\caption{Phonon frequencies $\omega_n$ at the zone center compared with
the experimental data of Refs. \onlinecite{phonons}, 
\onlinecite{Raman1}, \onlinecite{Raman2}, and \onlinecite{Raman3}
(all in meV). I and R denotes infrared and Raman active modes, 
respectively. The difference in phonon frequency due to local 
interactions $\delta\omega$ is given in percents.}
\begin{ruledtabular}
\begin{tabular}{c c c c c c}
          &        &\multicolumn{2}{c}{$\omega_n$}&      &            \cr
$\Gamma$ & active & $U=0$ & $U=4.0$ eV & $\delta\omega$ & Experiment \cr
\hline
$T_{2u}$  &    &  16.84  &  17.68 &  5.0 & $18.5^a$    \cr
$T_{1u}$  & I  &  19.98  &  21.46 &  7.4 & $12^b$      \cr
$E_u$     &    &  21.10  &  22.71 &  7.0 &             \cr
$T_{2g}$  & R  &  24.16  &  25.77 &  6.7 & $23.93^c$, $23.93^d$   \cr
$E_g$     & R  &  32.87  &  41.76 & 27.0 & $39.43^b$, $38.19^c$, $37.20^d$  \cr
$T_{1g}$  &    &  33.10  &  39.27 & 18.6 &             \cr
$A_{2u}$  &    &  35.53  &  38.08 &  7.2 &             \cr
$T_{1u}$  & I  &  38.31  &  40.10 &  4.7 & $32^b$      \cr
$T_{1u}$  & I  &  40.03  &  42.85 &  7.0 & $42.5^b$, $43.4^c$   \cr
$T_{2u}$  &    &  42.72  &  45.31 &  6.1 &   \cr
$T_{2g}$  & R  &  49.61  &  55.22 & 11.3 & $50.83^b$, $50.83^d$   \cr
$E_u$     &    &  52.50  &  54.30 &  3.4 &   \cr
$T_{2g}$  & R  &  65.10  &  68.89 &  5.8 & $67.20^b$, $66.95^c$, $66.95^d$  \cr
$T_{1u}$  & I  &  66.77  &  66.24 &  0.8 & $68^b$, $69.43^c$    \cr
$A_{1g}$  & R  &  73.06  &  82.74 & 13.2 & $83.32^b$, $83.07^c$, $82.95^d$  \cr
$A_{2u}$  &    &  74.21  &  81.33 &  9.6 &   \cr
\end{tabular}
\end{ruledtabular}
\leftline{$^a$ Reference \onlinecite{phonons}} 
\leftline{$^b$ Reference \onlinecite{Raman1}} 
\leftline{$^c$ Reference \onlinecite{Raman2}}
\leftline{$^d$ Reference \onlinecite{Raman3}} 
\end{table}

In this Section, we analyze the lattice dynamics of magnetite in the 
cubic phase. The low-energy dispersion curves were presented already in 
the previous paper.\cite{PRL} Here we extend this study by a detailed 
discussion of the phonon modes at the zone center and analyze the effects 
related to electron correlations. In the $\Gamma$ point there are 42 
modes classified according to the IRs of the cubic symmetry group,
\begin{equation}
\label{gamma}
\Gamma= A_{1g}+2A_{2u}+E_g+2E_u+T_{1g}+3T_{2g}+5T_{1u}+2T_{2u}.
\end{equation}
There are four optic infrared modes with $T_{1u}$ symmetry, five Raman 
modes with $A_{1g}$, $T_{2g}$, and $E_g$ symmetries, respectively, and 
seven silent modes. In Table VII we compare the frequencies of the optic 
modes with the experimental data from the neutron, infrared, and Raman 
measurements. Two sets of theoretical frequencies obtained from the 
calculations with: 
(i) $U=0$ and $J=0$ called below $U=0$, and 
(ii) $U=4.0$ eV and $J=0.8$ eV called below $U=4$ eV, are presented. 
The lowest optic mode at 18.5 meV with 
$T_{2u}$ symmetry (previously assigned as $T_{2g}$) was measured by the 
neutron scattering.\cite{phonons} Four infrared modes with $T_{1u}$ 
symmetry have been measured by Degiorgi {\it et al.} \cite{Raman1} 
Two lowest ones (at 12 and 32 meV) have very low intensities and 
large widths in experiment, so they can be hardly compared with the 
theoretical values. 
For the other two modes (at 42.5 and 67.2 meV) the agreement between 
the experiment and the present calculations is very good. In fact, in a 
broad range the frequencies $\omega_n$ of these modes are almost 
independent of $U$, but finite $U$ is of importance to improve the 
agreement with experiment for the mode observed at $\omega=42.5$ meV.

The theory shows that two frequencies around $40$ meV are very close to
each other, and it may be the reason of apparent difficulties in 
resolving these infrared modes. This interpretation could be supported 
by the fact that the peak observed in spectroscopy at $42.5$ meV seems
to be indeed very broad.\cite{Raman1} As a success of the present 
theory one finds that all computed Raman modes show good agreement with
experimental data. In two cases (for $A_{1g}$ and $E_g$ modes), 
calculations with finite $U=4.0$ eV largely improve this agreement. 
The only discrepancy found is the reversed assignement of symmetries 
($E_g$ and $T_{2g}$) of the two modes at $\sim 39$ and $\sim 50$ meV.
\cite{Raman2,Raman3} 
At present we cannot explain this discrepancy and we hope that future 
experiments would verify the present prediction of the theory.

Computed phonon frequencies show rather strong dependence on $U$. 
Interestingly, the frequencies {\it increase\/}, in spite
of larger lattice constant for $U>0$ which suggests the oposite trend. 
This indicates a strong effect of local Coulomb interactions on lattice
dynamics via respective changes in the electronic density distribution 
and in the value of EP coupling. It can be understood using a simple 
picture explained below. In the presence of local correlations, the 
electron localization in Fe($3d$) orbital states is enhanced. 
It results in weaker screening of ionic interactions and, consequently, 
in larger interatomic forces. Since phonon frequencies depend directly 
on force constants, their values increase. This mechanism is supported 
by the observation that the phonon modes stiffen in the insulating 
state in spite of lattice expansion due to the monoclinic distortion.
\cite{Raman2,Raman3,NIS1} One should note also that two Raman modes at 
frequencies $\sim 39$ meV ($E_g$) and $\sim 83$ meV ($A_{1g}$), which 
strongly depend on $U$, show the largest anomalies at the transition 
point.\cite{Raman2,Raman3} In contrast, for the infrared phonons the 
frequency dependence on $U$ is much weaker and no significant 
anomalies were found at $T_V$, apart from some changes in 
the linewidths of these modes.\cite{Raman2}

\begin{figure}[t!]
\includegraphics[width=7.7cm]{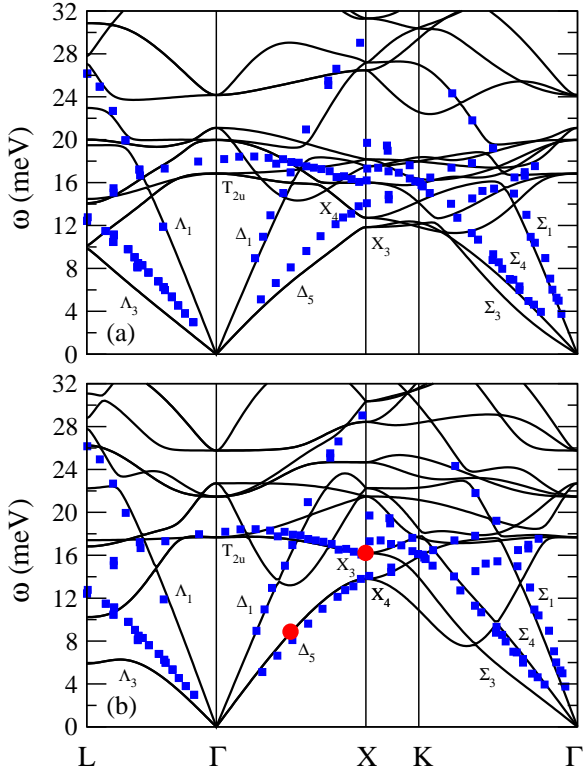}
\caption{(Color online)
Low-energy phonon frequencies $\omega$ as obtained for the cubic phase 
of Fe$_3$O$_4$ with:
(a) $U=J=0$, and 
(b) $U=4.0$ eV and $J=0.8$ eV. 
The squares show the experimental data obtained by neutron scattering.
\cite{phonons} Two primary OPs are related to $\Delta_5$ and $X_3$ 
phonons marked by circles in (b).
The high symmetry points from left to right in units of $\frac{2\pi}{a}$
are: $L=(\frac{1}{2},\frac{1}{2},\frac{1}{2})$, $\Gamma=(0,0,0)$,
$X=(0,0,1)$, $K=(\frac{1}{2},\frac{1}{2},1)$, $\Gamma=(1,1,1)$.
}
\label{fig:fig2}
\end{figure}

The phonon dispersion relations were calculated along the main directions 
of the reciprocal space. In Tab. VIII we present the 
compatibility relations for the IRs in the $\Gamma$ point and along 
three directions: $\Delta[001]$, $\Sigma[110]$, and $\Lambda[111]$. 
They show how the degenerate IRs split when symmetry is reduced. Knowing 
these relations, the symmetries of all dispersion curves can be properly 
assigned. Fig. \ref{fig:fig2} presents the lowest dispersion curves 
obtained for $U=0$ and $U=4.0$ eV, compared with the neutron scattering 
data.\cite{phonons} Longitudinal acoustic modes 
agree very well with experimental values, almost independently of the 
actual value of $U$ (the data for intermediate values of $U$ are not
shown). This behavior demonstrates that these modes couple very weakly 
to electronic density distribution, and are therefore not sensitive to
changes in electronic structure. A radically different behavior is 
observed for the transverse acoustic and optic phonons, which strongly 
depend on $U$. The most significant changes were obtained here in the 
$\Delta$ and $\Lambda$ directions. So far, we could not find a good 
physical explanation for a large discrepancy between theory and 
experiment observed at the $L$ point. 

\begin{table}[t!]
\label{tab:compa}
\caption{
Compatibility relations between the $\Gamma$ point and $\Delta$, 
$\Sigma$ and $\Lambda$  directions, 
as well as the $X$ point and $\Delta$, $\Sigma$ directions.}
\begin{ruledtabular}
\begin{tabular}{c c c c}
$\Gamma$, $X$   & $\Delta$ [001] & $\Sigma$ [110]  &  $\Lambda$ [111]  \cr
\hline
$A_{1g}$ & $\Delta_1$  & $\Sigma_1$ & $\Lambda_1$             \cr
$A_{2u}$ & $\Delta_4$  & $\Sigma_3$ & $\Lambda_1$             \cr
$E_{g}$ & $\Delta_1\oplus\Delta_2$ & $\Sigma_1\oplus\Sigma_3$ & $\Lambda_3$  \cr
$E_{u}$ & $\Delta_3\oplus\Delta_4$  & $\Sigma_2\oplus\Sigma_3$ & $\Lambda_3$ \cr
$T_{1g}$ & $\Delta_1\oplus\Delta_5$ & $\Sigma_1\oplus\Sigma_3\oplus\Sigma_4$ & $\Lambda_1\oplus\Lambda_3$  \cr
$T_{2g}$ & $\Delta_4\oplus\Delta_5$ & $\Sigma_1\oplus\Sigma_2\oplus\Sigma_3$ & $\Lambda_1\oplus\Lambda_3$  \cr
$T_{1u}$ & $\Delta_1\oplus\Delta_5$ & $\Sigma_1\oplus\Sigma_3\oplus\Sigma_4$ & $\Lambda_1\oplus\Lambda_3$  \cr
$T_{2u}$ & $\Delta_2\oplus\Delta_5$ & $\Sigma_1\oplus\Sigma_2\oplus\Sigma_4$ & $\Lambda_2\oplus\Lambda_3$  \cr
\hline
$X_{1}$ & $\Delta_1\oplus\Delta_4$ & $\Sigma_1\oplus\Sigma_3$ \cr
$X_{2}$ & $\Delta_2\oplus\Delta_3$ & $\Sigma_2\oplus\Sigma_4$ \cr
$X_{3}$ & $\Delta_5$  & $\Sigma_3\oplus\Sigma_4$ \cr
$X_{4}$ & $\Delta_5$  & $\Sigma_1\oplus\Sigma_2$ \cr
\end{tabular}
\end{ruledtabular}
\end{table}

For the lowest acoustic and optic $\Delta_5$ modes, the agreement 
improves in case of $U=4$ eV. It is quite remarkable that the optic 
mode behaves differently in both cases. For $U=0$ this mode crosses 
with another $\Delta_5$ mode, so the sequence of phonon branches is 
interchanged. This behavior is responsible for the apparent 
discrepancies between the frequencies of phonon modes in theory and 
experiment for the $U=0$ case. According to the experiment,
\cite{phonons} the lowest phonon at the $X$ point has the $X_4$ 
symmetry, and splits into the $\Sigma_1$ and $\Sigma_2$ phonon 
branches along the $[110]$ direction (see Table VIII).
The second lowest mode has $X_3$ symmetry and transforms into the 
$\Sigma_3$ and $\Sigma_4$ modes. This order is indeed accurately 
reproduced by the GGA+$U$ calculations with realistic local Coulomb 
interactions (i.e., taking $U=4.0$ eV and $J=0.8$ eV). In contrast, 
in the $U=0$ case, the order of the lowest modes is reversed. It shows 
that Coulomb interactions strongly modify lattice dynamics and the 
phonon energy spectrum. We thus emphasize that they have to be included
to obtain not only quantitative but even qualitative agreement with 
experiment. 

\section{Electron-phonon interaction}

\subsection{Instability of the cubic phase}
\label{sec:insta}

Having obtained phonon dispersion curves and their symmetries, we now 
focus on the effects induced by phonons, especially on the OPs derived 
in Sec. II. In this Section, we discuss the EP interactions associated 
with the $X$ and $\Delta$ phonons, and present the instability of the
cubic phase triggered by local deformations. Since one expects the 
strongest effects for the low energy phonons, we analyze only the lowest
energy acoustic and optic branches. 

\begin{figure}[t!]
\includegraphics[width=7.7cm]{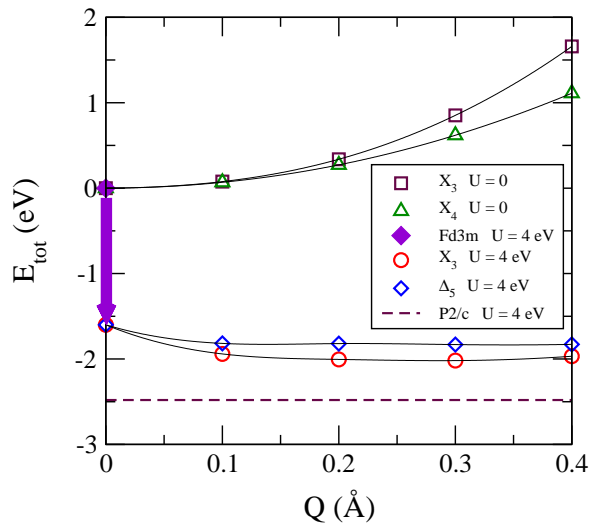}
\caption{(Color online) 
Total energy $E_{\rm tot}$ (\ref{etot}) as a function of the amplitude 
$Q$ of phonon modes with the $X_3$, $X_4$, and $\Delta_5$ symmetries 
(in \AA{}), as obtained with ($U=4$ eV) and without ($U=0$) local Coulomb 
interactions at Fe ions. The energy gain of $\sim 1.6$ eV due to the 
orbital polarization at $Q=0$ is indicated by the arrow.
}
\label{fig:fig3}
\end{figure}

First, for $\textbf{k}_X$ wave 
vector we compare the EP coupling for the lowest $X_3$ and $X_4$ modes. 
At the $\textbf{k}_X$ point all modes are doubly degenerate, so we used 
polarization vectors from only one branch by shifting the wavevector to
$\textbf{k}_X+\delta\textbf{k}_{\Sigma}$, removing then the degeneracy.
In Fig. \ref{fig:fig3} we present the total energy $E_{\rm tot}$ of the 
distorted cubic structure as a function of the increasing phonon 
amplitude $Q$. The energies of relaxed (undistorted) crystals obtained in 
the GGA and GGA+$U$ calculations were shifted to a common energy origin 
for a more transparent presentation (their actual values can be found 
in Table IV). One finds that for $U=0$ the energy increases when the 
cubic structure is distorted by either $X_3$ or $X_4$ phonon. This 
behavior is characteristic and represents a typical situation when atoms 
are displaced from their respective equilibrium positions. It simply 
confirms that the cubic symmetry is the most stable structure for $U=0$,
and does not undergo any instability towards other possible symmetries. 

In the GGA+$U$ calculations, the situation changes drastically since the 
orbital degrees of freedom becomes active and influence the electronic 
structure. As demonstrated in Ref. \onlinecite{JT2}, the orbital polarization 
breaks the symmetry of wavefunctions within the cubic $Fd\bar{3}m$ structure 
and leads to lowering of the total energy. This effect was demonstrated by 
initializing the self-consistent procedure using the wavefunctions of the 
$C2/C$ structure.\cite{JT2} Applying a similar procedure, we have started 
calculations with the wavefunctions of the low-symmetry $P2/c$ phase and 
optimized the electronic structure self-consistently in the $Fd\bar{3}m$ 
structure. The reference energy obtained by the direct optimization is 
denoted by the point at $E_{\rm tot}=0$ in Fig. \ref{fig:fig3}, and the 
decrease due to orbital polarization (about 1.6 eV) is visualised by the 
arrow. The lowering of the ground state energy is associated with a partial 
CO and OO which arises in the $t_{2g}$ states, however, it does not 
generate the gap opening yet. In fact, the calculation suggests that 
this is rather a metastable state, showing only short-range charge-orbital 
correlations without well defined symmetry. 

Total energy $E_{\rm tot}$ {\it decreases\/} further from that obtained 
for the state with CO and OO induced by finite Coulomb interactions for 
increasing phonon distortion $Q$ (see Fig. \ref{fig:fig3}), which leads 
eventually to the metal-insulator transition. The strongest decrease of 
the total energy was found for the $X_3$ mode. On the contrary, by 
considering a single $X_4$ phonon amplitude we have verified that this 
type of distortion does not induce any energy decrease. Instead, the 
acoustic $\Delta_5$ phonon mode leads to the energy lowering when local 
Coulomb interactions are present. Therefore, we included in Fig. 
\ref{fig:fig3} also the energy dependence for this acoustic mode. The 
energy decrease is here smaller than in  case of the $X_3$ mode. Note 
however that because of double degeneracy the total energy may also 
depend on the phase. Finally, the groundstate energy for the $P2/c$ 
symmetry is shown in Fig. \ref{fig:fig3} as the lowest dashed line. 
Since the lattice deformation induced by either $X_3$ or $\Delta_5$ 
mode is not sufficient to lower the energy to this level, other OPs 
have to be active as well.

\subsection{Metal-insulator transition}
\label{sec:mit}

\begin{figure*}[t!]
\includegraphics[width=16cm]{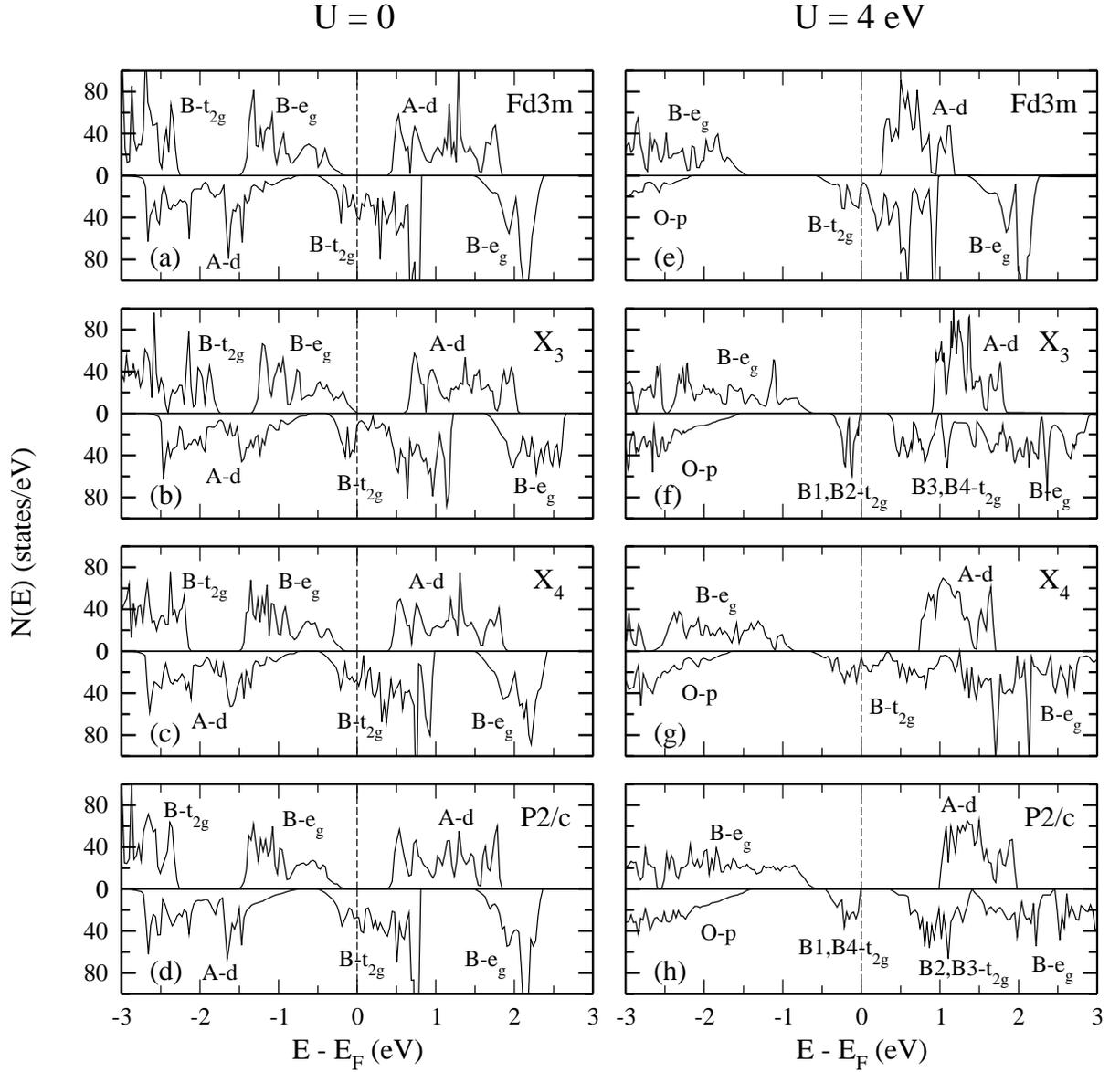}
\caption{
Electronic DOS for the spin-up (top) and spin-down (bottom) electrons 
in the phases of $Fd\bar{3}m$, $Fd\bar{3}m+X_3$, $Fd\bar{3}m+X_4$, 
and $P2/c$ symmetry, as obtained in: 
(a)--(d) GGA ($U=0$), and 
(e)--(h) GGA+$U$ (with $U=4.0$ eV) calculations.}
\label{fig:fig4}
\end{figure*}

The changes in the electron DOS induced by the considered phonon modes 
at the $\textbf{k}_X$ point are presented in Fig. \ref{fig:fig4}. 
They are compared with the respective DOS found for the undistorted 
structures with $Fd\bar{3}m$ and $P2/c$ symmetry, respectively. The 
results of the calculations performed in the GGA and in the GGA+$U$ 
approach are shown on the left and right hand side of Fig. 
\ref{fig:fig4}, respectively. The main effect of finite $U$ for the 
cubic structure is to increase the exchange splitting between the up 
and down spin states, cf. Figs. \ref{fig:fig4}(a) and 
\ref{fig:fig4}(e). The up-spin states at Fe($B$) ions and down-spin 
states at Fe($A$) ones are shifted to lower energies below $E_F$. 
In addition, finite $U$ decreases the spectral density of minority 
$t_{2g}$ states just above $E_F$. This indicates the enhancement of 
electron localization due to local electron interactions, which leads 
to gap opening in the distorted structure.

First observation connected with the EP interaction is that the overall 
effects of $X_3$ and $X_4$ distortion are much weaker for the 
uncorrelated ($U=0$) case than for the correlated ($U=4.0$ eV) one. 
Comparing these two distortions for $U=0$ [Figs. \ref{fig:fig4}(b) and 
\ref{fig:fig4}(c)], we see that the strongest coupling is associated 
with the $X_3$ phonon. Especially for down-spin states above $E_F$ one 
finds a decrease of the spectral density. This instability is connected 
with the $\textbf{k}_X$ nesting vector at the Fermi surface, as 
discussed some time ago by Yanase and Hamada.\cite{DFT3} 
It is quite remarkable that this effect alone does not suffice to 
induce a metal-insulator transition, but such a transition is triggered 
when it is amplified by local Hubbard interaction $U$. Indeed, when the 
$X_3$ distortion is made in presence of finite $U$ 
[Fig. \ref{fig:fig4}(f)], the gap opens at $E_F$, and the 
metal-insulator transition takes place. Note that apart from some 
subtle differences, the distribution of the spectral weight near $E_F$ 
found with the $X_3$ distortion is quite similar to that found in the
$P2/c$ phase [Fig. \ref{fig:fig4}(h)].

Surprisingly, the $X_4$ mode does not produce a similar effect, 
although the changes of the electron DOS were here similar to those
due to $X_3$ for $U=0$, and the EP coupling is here also much stronger 
than for $U=0$. This result agrees perfectly with the group theory 
analysis of Sec. II, showing that $X_3$ is the primary OP and $X_4$ is 
neither primary nor secondary OP for the VT. We remark that the 
influence of the $\Delta_5$ mode on the DOS has been discussed 
previously,\cite{PRL} and it does not induce the insulating gap either
in spite of its significant coupling to the electronic density. 
Thus, we conclude that the $X_3$ mode plays the decisive role in the 
metal-insulator transition which accompanies the VT.

\subsection{Charge and orbital order}
\label{sec:oo}

\begin{figure}[t!]
\includegraphics[width=8.2cm]{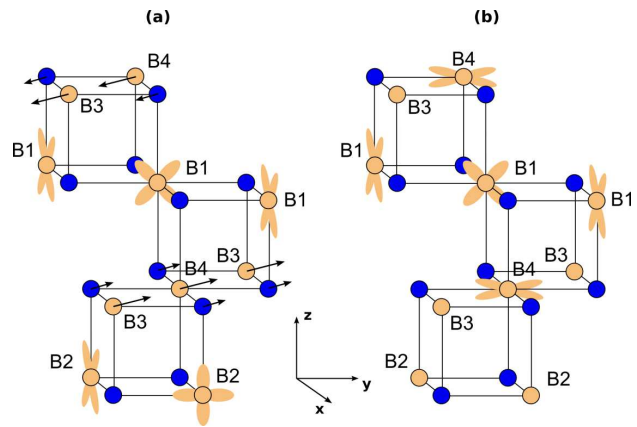}
\caption{(Color online) 
Orbital order in the $t_{2g}$ states of Fe($B$) ions in magnetite as 
found for:
(a) the $Fd3m$ cubic structure distorted by the $X_3$ phonon, and 
(b) the structure of $P2/c$ symmetry. 
The balls indicate Fe and O ions in the respective crystal structure, 
with the same meaning of colors as in Fig. \ref{fig:fig1}.
The arrows in (a) represent the atomic displacements in the $X_3$ mode.}
\label{fig:fig5}
\end{figure}

Realistic treatment of the charge and orbital order requires the
electronic structure calculations with finite local Coulomb 
interactions using LSDA+$U$-like methods.
The electronic DOS in the cubic phase [Fig. \ref{fig:fig4}(e)], as well as 
earlier experimental studies,\cite{Gar00} suggest that magnetite is an 
itinerant magnet and not a mixed valent system. The VT is therefore not
due to freezing of charge fluctuations at Fe($B$) ions, but rather due 
to the symmetry change, which leads only to weak CO as a consequence of 
the undergoing phase transition. This scenario is confirmed by the 
changes in the electron DOS above and below the VT, as the gap at the 
Fermi energy induced by the $X_3$ phonon [see Fig. \ref{fig:fig4}(f)] 
splits the $t_{2g}$ band into the occupied states at the $B1$ and $B2$ 
Fe ions and empty states at the $B3$ and $B4$ ions. Thus, this mode is 
responsible for the charge disproportionation which persists also in 
the $P2/c$ phase [Fig. \ref{fig:fig4}(h)]: increased $t_{2g}$ electron 
density for $\{B1,B2\}$ ions and reduced density for $\{B3,B4\}$ ions. 
Note that four Fe($B$) sites split into two subclasses as a result of 
the lattice distortion in the $X_3$ mode [Fig. \ref{fig:fig5}(a)]. The 
amplitude of the charge (spin) disproportionation is comparable with 
that found for the $P2/c$ symmetry (see Table VI). Interestingly, this 
CO pattern is {\it similar\/} to that proposed originally by Verwey and 
fulfills the Anderson criterion. These findings agree also with other 
LSDA+$U$ and GGA+$U$ calculations.\cite{prl1,Leonov,JT2}

In addition, we have found that the $X_3$ mode stabilizes the OO in the 
$t_{2g}$ states. The resulting orbital polarization with atomic 
displacements is presented in Fig. \ref{fig:fig5}(a). In the considered 
mode the atoms vibrate along the $[110]$ direction, and the dominating
components are those connected with distortions of Fe($B$) ions and 
oxygen ions in the same planes. Oxygen displacements are 65\% of 
Fe($B$) displacements, while Fe($A$) atoms shift only by about 23\% of 
them (not shown). The considered mode is limited to single Fe--O 
planes: Fe($B$) ions move along with O ions in one plane, while atoms 
in neighboring planes do not move at all (they participate in the other 
$X_3$ branch in the perpendicular direction). This deformation modifies 
distances between Fe($B$) atoms and oxygens situated above and below 
them, inducing changes in the electron density distribution within 
$t_{2g}$ states and a coexisting charge and orbital order in the $B1$ 
and $B2$ chains. As indicated in Fig. \ref{fig:fig5}, electrons occupy 
orthogonal $d_{yz}$ and $d_{xz}$ orbitals, thus forming the state with 
alternating OO. 

This coexisting charge-orbital order and the insulating gap induced by 
the $X_3$ mode has to be compared with that observed in the monoclinic 
phase of $P2/c$ symmetry [Fig. \ref{fig:fig5}(b)]. The electronic DOS 
for this symmetry is presented in Fig. \ref{fig:fig4}(h). The magnitude 
of the gap is similar to the gap generated by the $X_3$ phonon mode 
shown in Fig. \ref{fig:fig4}(f), however, the occupied states below 
$E_F$ belong now to the $\{B1,B4\}$ ions and empty states to the 
$\{B2,B3\}$ ones. Actually, Fe($B$) ions are split into two groups 
$\{B1,B4\}$ and $\{B2,B3\}$, as average Fe--O distances for the bonds 
to the $\{B2,B3\}$ sites are significantly smaller than those for the 
$\{B1,B4\}$ ones. Therefore, the CO has to change from that promoted by 
the $X_3$ mode alone. Surprisingly, the weak CO found in the LT 
monoclinic phase does not satisfy the Anderson criterion, but plays 
no role in the actual mechanism of the transition.

The corresponding orbital pattern also has changed [see Fig. 
\ref{fig:fig5}(b)]. The occupied states on $B1$ and $B4$ irons form the 
alternating OO with the orthogonal orbitals on the nearest neighbor 
iron sites. We emphasize that in the $P2/c$ frame, the occupied 
$t_{2g}$ states are given by $\frac{1}{\sqrt{2}}(d_{xz}\pm d_{yz})$ and 
$d_{x^2-y^2}$ combinations. 
Therefore, on comparing Figs. \ref{fig:fig5}(a) and \ref{fig:fig5}(b), 
we conclude that the $X_3$ mode demonstrates a generic tendency 
towards the OO within the $t_{2g}$ minority states, but only partly 
explains the observed OO in the LT phase with $P2/c$ symmetry. This is 
however quite natural and expected, since the $X_3$ mode is only one 
component of the OPs that condense at the VT, which {\em all\/} finally
lead to the change of symmetry to the LT monoclinic phase.

In the next step, one should consider the superposition of the $X_3$ 
with the $\Delta_5$ mode. The latter one is crucial to explain the 
doubling of the unit cell and the observed charge-orbital pattern.
As we mentioned before, this mode is double degenerate and the atomic
displacements depend on the phase. So, a detailed analysis of the 
EP coupling, similar to that presented above for the $X_3$ modes, is 
rather difficult. Moreover, because of the system complexity, it is 
likely that the atomic displacements belonging to the secondary OPs 
are also involved in the final charge-orbital order. In spite of these 
difficulties, the studies presented in this Section demonstrate how 
the phonons interplay with charge and orbital degrees of freedom in 
the VT.

\section{Mechanism of the Verwey transition}

From the results presented in the previous Section and from the earlier 
experimental studies, it is clear that the mechanism of the VT includes 
two essential ingredients: 
(i) strong intraatomic Coulomb interactions $\{U,J\}$ at Fe ions, and 
(ii) phonon-driven lattice instability.
On the one hand, the electron interactions are responsible for the 
orbital polarization, which breaks the symmetry of the wavefunction and 
enhances the electron localization in the $t_{2g}$ states. On the other
hand, the phonon OPs induce the crystal distortion, which generates the 
structural transformation from the cubic $Fd\bar{3}m$ phase to the LT 
phase with monoclinic $P2/c$ symmetry. The coupling between these two 
subsystems is crucial since none of them alone would be able to explain 
the metal-insulator transition which occurs in magnetite at $T_V$. As 
we have shown, without the Hubbard interaction $U$, the EP interaction 
would be too weak to induce structural changes. In the presence of 
electron correlations, the coupling between the electrons in $t_{2g}$ 
states and phonons is largely enhanced since it leads to stabilization 
of the OO and to lowering of the total energy. The corresponding CO is 
very subtle and seems to be rather a consequence of the joint effect of 
the OO and enhanced EP interactions than the driving force of the VT.

The splitting of the $t_{2g}$ states due to the EP interaction is 
usually referred to as the Jahn-Teller effect. In principle, the 
mechanism observed in magnetite is similar to this effect, however, it 
differs in many respects from the classical examples. In a typical 
situation, one considers the displacements of oxygen ions around the 
transition metal cations, which break the symmetry and split the 
degenerate $3d$ states. As we discussed in the previous Sections, the 
displacement pattern in the monoclinic structure is more complicated 
and involves also $Fe($B$)$ ions. In particular, in the $X_3$ mode the 
iron atoms (with largest amplitudes) vibrate in the same plane as 
oxygens. Therefore, the orbital polarization results in this case 
mainly from the interplanar interactions between iron and oxygen ions.

A dynamical coupling between the charge-orbital fluctuations and 
phonons induces the critical diffuse scattering above $T_V$. 
Temperature dependence of this scattering gives us important 
information about precursors, which show up far from the critical 
point. The signs of the transition appear already at about 200 K. 
Apart from neutron scattering coming from ${\bf k}_{\Gamma}$ and 
${\bf k}_X$ points,\cite{diff3,diff5} there are maxima at other 
positions, also with an incommensurate wavelengths. For instance, very 
pronounced critical behavior was observed at ${\bf k}=(8,0,0.75)$.
\cite{diff2} When temperature is lowered, the coupling between 
different modes, described by the Landau free energy (\ref{free}), 
leads to stabilization of particular phonons with commensurate 
wavelengths. Very close to the transition point, about 5 K above $T_V$,
the $\Delta_5$ OP becomes active, producing the signal in neutron 
measurement.\cite{diff1} The question arises why is this mode observed 
only a few kelvins above $T_V$, while signals from the $\Gamma$ and $X$ 
points are seen at much higher temperatures? This may be connected with 
the relative strength of different OPs. In Fig. \ref{fig:fig2}, we see 
that the coupling to $\Delta_5$ mode is weaker than to $X_3$ mode, and 
it gives a different signal in neutron scattering.

Critical scattering is one of indications of a short-range (charge and
orbital) order above the VT, first discussed by Anderson.\cite{anderson} 
Another evidence is connected with the change in entropy, which is 
smaller than in the order-disorder phase transitions.\cite{entropy} 
Also the temperature dependence of resistivity at $T>T_V$, which is not 
typical for metals, suggests the polaronic nature of charge carriers.
First photoemision measurements indicated the gap closing above $T_V$, 
\cite{PES1} but more detailed studies\cite{PES2,PES3,PES4} established 
that the gap only decreases, without any sharp change at $T_V$. The 
optical conductivity studies suggest also the gap opening below $T_V$, 
but above the transition the conductivity spectrum does not exhibit a 
metallic Drude-type behavior.\cite{OC} Instead, a hopping type 
conductivity is observed, with highly diffuse character of charge 
dynamics. This picture is supported by the structural EXAFS studies, 
which found that the local crystal geometry does not change at $T_V$.
\cite{EXAFS} It means that the quasistatic lattice distortions are 
present already in the high-symmetry cubic phase above $T_V$. All these 
results are consistent with the present theory, which shows that some 
phonons strongly couple to electronic states and may induce local 
crystal deformations and polaronic short range order above $T_V$.

As demonstrated in this work and the previous study,\cite{PRL} the 
theoretical approach based on the GGA+$U$ (or LDA+$U$) provides a very 
good description of the VT. However, there are some aspects, which 
involve dynamical processes and many-body interactions. For example, 
the detailed analysis of the photoemision spectra showed that the DFT 
is not sufficient to explain changes induced by the VT and better 
agreement with experiment can be achieved only by using the DMFT 
approach.\cite{JT1} An interesting effect was observed in the spin 
excitation spectrum, which shows a large splitting in the acoustic 
magnon branch at $\textbf{k}_{\Delta}$ below the VT.\cite{spinphonon1}
This effect cannot be explained merely by the dependence of exchange 
interactions on the crystal structure or charge ordering, but is 
rather a consequence of the magnon-phonon coupling. 
Additionally, a recent neutron scattering study revealed
an anomalous broadening and energy shift of the $\Delta_5$
spin wave above $T_V$.\cite{spinphonon2}
A strong spin-phonon interaction, suggested by the present work, provides 
a sound starting point for studying such effects.

Finally, the behavior of magnetite at high pressure is rather intriguing.
According to diffraction studies,\cite{roz1} the critical temperature $T_V$ 
decreases with pressure, so the VT can be induced at rather low 
temperature by applying pressure in the LT monoclinic phase.
Recently two contradicting views on the VT based on 
pressure experiments have been suggested. On the one hand, a possible 
transition from the inverse to normal spinel structure 
has been suggested as consistent with the interpretation of the 
M\"{o}ssbauer and x-ray scattering measurements.\cite{pasternak,roz2} 
This scenario precludes a CO on the Fe($B$) sites below $T_V$, but 
predicts a large $\sim 50$\% increase in the bulk magnetization with 
increasing pressure or by lowering temperature. 
On the other hand, no change in magnetic moments at both $A$ and $B$
sites has been detected by the neutron diffraction studies in a broad 
range of pressure up to $p=5.3$ GPa.\cite{klotz} This latter result 
rules out a posibility of the inverse spinel to normal spinel
transition under decreasing temperature (increasing pressure), 
at least in the regime of pressure lower than $5.3$ GPa.
It also agrees with earlier band structure calculations,\cite{DFT2} 
and with the present mechanism of structural transition in magnetite. 
As the electron occupations at $A$ and $B$ sites are modified rather 
weakly, no significant change in the values of magnetic local 
moments is expected. We emphasize that this latter view does not imply 
large CO below the VT. 
In fact, the structural transition suggested in the present work not 
only {\it explains\/} the dramatic change in the conductivity observed 
at $T_V$, but also {\it predicts\/} rather weak CO, in agreement with 
experiment.\cite{CO3,CO4,CO5,CO6}

\section{Summary and conclusions}

In this work, we have presented a detailed group theory analysis
of the Verwey transition. We have identified three primary OPs: $X_3$, 
$\Delta_5$, and $T_{2g}$, which describe the symmetry reduction in the 
crystal structure transformation from the high-temperature cubic $Fd3m$ 
phase to the LT monoclinic $P2/c$ phase. By performing the numerical 
{\it ab initio\/} computations, we have demonstrated that a prominent
role is played by the $X_3$ mode which: 
(i) couples strongly to the electronic states, 
(ii) lowers the total energy, and 
(iii) is responsible for the metal-insulator transition. 
The latter transition occurs only when local Coulomb interaction $U$ is 
explicitly included in $t_{2g}$ iron states and may trigger weak charge 
order at $B$ sites. In fact, this pnenomenon appears counterintuitive 
and occurs only as a result of the accompanying local lattice 
distortions which may be seen as a Jahn-Teller lattice instability. 
These results support the recent point of view\cite{PRL,Leo07} that 
neither pure electrostatics is the main factor responsible for the 
charge order observed below the Verwey transition, nor the charge order 
is the mechanism driving the transition by itself. At the same time, 
the local Coulomb interaction generates alternating orbital order, 
which leads to strong reduction of charge mobility and amplifies 
electron-lattice effects.

The present study reconciles several previous points of view on the 
Verwey transition in magnetite and suggests that the physical effects 
which occur simultaneously below this transition in the monoclinic 
$P2/c$ phase can be classified into the ones which are the primary 
cause of the symmetry change and the ones which occur only as its 
consequence. In this way it contributes to the recent debate concerning
the origin of the transition and clarifies the role played in it by the 
charge order. While weak charge order has been found at Fe($B$) sites, 
it is not surprising that it does not obey the Verwey model. 
In fact, it is only one of the manifestations of strong local electron 
interactions in partly occupied $t_{2g}$ states rather than the primary 
cause of the observed symmetry change.

We have also compared phonon energy spectrum of magnetite with the 
experimental data obtained by Raman and infrared spectroscopy, as well 
as by neutron scattering. We have found that phonon frequencies 
strongly depend on local electron interactions, and the GGA+$U$ 
calculations performed with realistic parameters for Fe ions ($U=4.0$ 
eV and $J=0.8$ eV) give very satisfactory qualitative and quantitative 
agreement with the experimental data. In contrast, when the electron 
interaction effects are neglected, the phonon spectra are even 
qualitatively different from the observed ones. 

Summarizing, we have shown that the Verwey transition is promoted by
a set of order parameters with mixed electron and phonon character. 
The insulating monoclinic $P2/c$ phase occurs below the transition as 
a result of the instability driven by the electron-phonon coupling in 
presence of strong electron correlations. We argue that the 
electron-lattice coupling plays an important role also in other
transition metal oxides with strongly correlated electrons. Further
studies of the lattice relaxation effects in these systems may lead 
to discoveries of new electronic phenomena that could be understood
only by simultaneous treatment of electronic and lattice degrees of
freedom.

\acknowledgments
The authors thank A. Koz\l{}owski, J. \L{}a\.{z}ewski, and P.T. Jochym 
for valuable discussions.
This work was partially supported by Marie Curie Research Training Network
under Contract No. MRTN-CT-2006-035957 (c2c).
A.~M.~Ole\'s would like to acknowledge support by the Polish Ministry
of Science and Education under Project No. N202 068 32/1481.

\end{document}